\documentclass[11pt]{article}
\usepackage{moriond,epsfig}

\def\be{\begin{equation}}
\def\ee{\end{equation}}
\def\bea{\begin{eqnarray}}
\def\eea{\end{eqnarray}}

\begin{document}
\vspace*{4cm}
\title{Possible Dark Matter Signals from Antiprotons, Positrons, X-rays and Gamma-rays}

\author{U. Schwanke}

\address{Humboldt University Berlin, Department of Physics, Newtonstrasse 15,\\
12489~Berlin, Germany}

\maketitle\abstracts{
A number of signals involving charged cosmic rays and high-energy photons 
have been interpreted as being due to annihilating dark 
matter. This article provides an overview of the experimental evidence 
and discusses in particular detections of antiprotons and positrons
in the cosmic radiation, the diffuse $\gamma$-ray emission between 10\,MeV and 100\,GeV
from the Milky Way, and the 511\,keV annihilation radiation and 
the flux of very high-energy photons ($>100$\,GeV) from the Galactic Centre.
}

\section{Introduction}

Observations of numerous astronomical systems (rotation curves of galaxies, 
dynamics of galaxy clusters) have shown that the amount of gravitationally bound mass is much larger than the 
mass inferred from measurements in various wavelength bands.
Unless one is willing to modify the law of gravity at scales larger than 
the size of the solar system, one is forced to the conclusion that those
systems contain a substantial fraction of matter that contributes to gravity but does not shine.
Further compelling evidence for the existence of dark matter has been derived
from the high-precision WMAP measurements of the temperature 
fluctuations of the cosmic microwave background radiation. Apparently,
matter accounts for 27\,\% of the total mass density of the Universe (the remaining
73\,\% being dark energy), but the contribution of baryonic matter is only 4\,\%.
This implies that some form of non-baryonic dark matter dominates the
matter contribution to the total mass density of the Universe.

\section{Direct and Indirect Dark Matter Searches}

Particles that could constitute dark matter must be sufficiently heavy
and interact so weakly with ordinary matter that only their gravitational 
interaction has come to our attention up to now. None of the known particles 
has the right properties (neutrinos being too light for example), but there
is no lack of candidates proposed within various theoretical frameworks\,\cite{review}.

From an experimental point of view, searches at collider experiments
have yielded limits on the mass and the couplings of suitable particles 
that are predicted for example by extensions of the Standard Model of elementary
particle physics (like supersymmetry). Other searches do not seek
to produce dark matter particles with same man-made apparatus, and are then classified
as either {\em direct} or {\em indirect}. Direct searches aim at dark matter
particles that might be present in Earth's immediate neighbourhood and try to make them interact in
suitable detectors. The used detectors record recoil and
ionization signals that might be caused by scattering of dark matter particles
with the target material. Direct search experiments have yielded upper
limits on the interaction cross section of dark matter particles with
atoms, often as a function of the mass of the particle\,\footnote{There is 
one signal claim by the DAMA collaboration, but other experiments already excluded the
cross section region suggested by this measurement.}. Indirect searches
do not target dark matter particles directly, but rather their annihilation
or decay products and record these possibly far away from the location where
the interaction occured. In the following, we will concentrate exclusively
on indirect searches, discuss the basic principles and review 
signals that have been interpreted as possible evidence for dark matter.

\section{Indirect Searches}

Indirect searches detect secondaries arriving in Earth's immediate neighbourhood.
They can potentially benefit from cumulative effects from large amounts of 
dark matter (e.g.\ in the entire Milky Way) or from 
dark matter concentrations higher than in the solar system (e.g.\ in 
the centre of the Milky Way). On the other hand,
indirect searches offer a much less controlled experimental environment than direct searches,
and the interpretation of measurements requires knowledge of the 
dark matter distribution and a good understanding of less exotic
processes that might create the same secondaries.

Reasonable candidates for observable secondaries are 
antiprotons and positrons since those can be readily identified in the flux of 
ordinary charged cosmic rays. Neutrinos and $\gamma$-rays can provide
usable signals as well, and we will briefly discuss each particle type with 
its specific signature and related problems.
\begin{description}
\item[Antiprotons:] The signature expected from antiprotons
  from dark matter annihilation or decay is an energy spectrum 
  with a cutoff at the mass of the dark matter particle. Like all
  charged particles, (anti-)protons created somewhere in the Galaxy 
  quickly lose their directional information due to deflections 
  in the galactic magnetic field and spread out in a diffusive way. 
  The range of antiprotons   
  in the interstellar medium is high, so potentially the
  dark matter in the entire Milky Way can contribute to an antiproton
  signal observed on Earth. The major background are antiprotons 
  produced in interactions of cosmic ray protons with the 
  interstellar medium (mostly neutral and ionized hydrogen) 
  or even in Earth's atmosphere. Another complication arises
  from the fact that the flux of charged particles arriving on
  Earth varies not only with latitude (due to Earth's magnetic
  field) but also with time due to the Sun's activity. This
  so-called solar modulation can be neglected at energies
  above 10\,GeV, but is very important around 1\,GeV.
  
  The background subtraction and interpretation of 
  antiproton signals requires therefore a modelling of the Milky Way
  and an accurate description of propagation effects. Current  
  models\,\cite{model1,model2} treat particle diffusion, energy losses
  and spallation processes, but their reliability is hampered
  by the fact that some inputs (e.g.\ the distribution of
  cosmic ray sources and hydrogen in the galaxy, the energy dependence of diffusion
  coefficents etc.) have not been measured with high precision.
\item[Positrons:] Much of what has been said about antiprotons also
  applies to positrons. The biggest difference is that positrons 
  suffer from larger energy losses which is restricting their diffuse 
  range to a few kiloparsecs (kpc). This range is much smaller than the diameter
  of the Milky Way ($\approx 30\,\mbox{kpc}$), which implies that
  only dark matter within a few kpc would contribute to a signal detectable
  on Earth. On the bright side, this restriction to the solar neighbourhood  
  makes the treatment of the positron diffusion easier and reduces
  the impact of uncertainties of the dark matter distribution in the 
  galaxy (where $N$-body simulations predict a steep increase of
  the dark matter density towards the Galactic Centre). Background
  positrons are created in interactions of photons and charged cosmic
  rays, but there is also a number of astrophysical positron 
  sources (supernova explosions, pulsars etc.).
\item[$\gamma$-rays:]
  In contrast to charged secondaries, the measured arrival direction
  of photons can be correlated with the (dark) matter density. A clear
  dark matter signature would be the observation of $\gamma$-ray 
  lines from annihilation of dark matter particles at rest. In supersymmetric 
  theories, the final states of these annihilations ($\gamma\gamma$ or $\gamma Z$) can
  only be reached in loop diagrams and are therefore suppressed. For
  photons created in subsequent decays of annihilation products one
  expects an energy spectrum with a strict cutoff given by the mass of 
  the annihilating dark matter particle. Prominent backgrounds are created
  in interactions of charged cosmic rays with the interstellar
  medium (i.e.\ via $\pi^0$ decay), by bremsstrahlung and by inverse
  Compton scattering of electrons.
\item[Neutrinos:]
  Neutrinos are great messenger particles pointing back to their sources 
  and with hardly any propagation effects, but they are also difficult to detect. 
  Attempts to measure the flux of neutrinos created by dark matter 
  particles accumulating in the Sun or the Earth have resulted in upper
  limits that are not very constraining. We will not be concerned with neutrinos here.
\end{description}

\section{Antiprotons and Positrons}

\begin{figure}
\begin{tabular}{cc}
\epsfig{figure=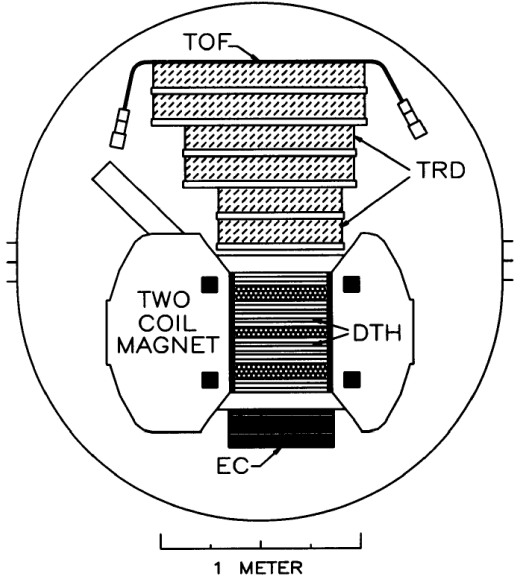,width=0.5\linewidth} &
\epsfig{figure=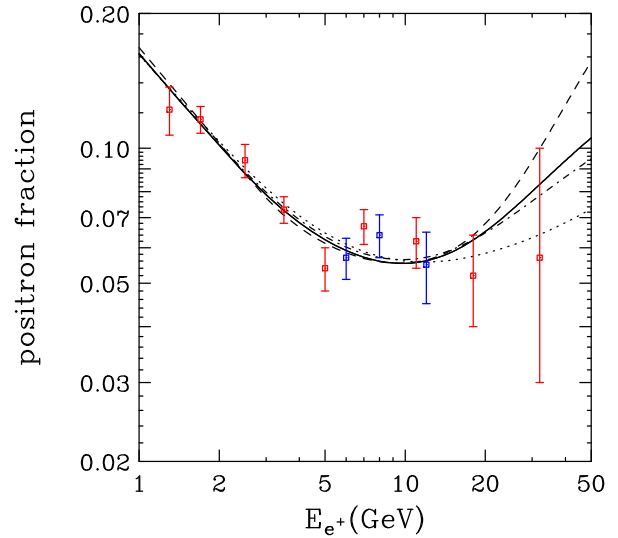,width=0.5\linewidth} 
\end{tabular}
\caption{\label{fig:heat} (Left:) Schematic view of the HEAT-$e^\pm$ spectrometer. 
         (Right:) Positron fraction measured by HEAT in 1994/95 (red) and 2000 (blue).
         The lines indicate a contribution from annihilating Kaluza-Klein dark matter 
         which has been normalized to the data. The various line styles correspond
         to different choices for the positron diffusion parameters\,\protect\cite{heathooper}.
}
\end{figure}

The detection of antiprotons and positrons in the upper atmosphere employs 
particle spectro\-meters (Fig.~\ref{fig:heat} (left)) carried by balloons to heights of around 40\,km
above sea level. The reached exposure is somewhat limited since these experiments
(like BESS or HEAT)
are flown typically once a year for a duration of roughly ${\cal O}(10)$ hours. 
The BESS experiment\,\cite{bessmitchell}, for example, recorded around 2000 antiprotons
in eight flights which took place in northern Canada over a period of 
11 years. The small fraction of antimatter in the cosmic radiation ($\bar{p}/p\approx 10^{-5}$)
poses a particular challenge to the balloon-borne spectrometers, and the large
background of particles with the wrong charge (i.e.\ protons and electrons)
requires high precision tracking and excellent particle identification 
capabilities. Some of the associated experimental uncertainties were 
maybe not so well known in early experiments whose data suggested a 
much larger number of antiprotons and positrons than expected from purely secondary 
production. Meanwhile, both the understanding of the measurements and of the 
expected background of secondary antiprotons have improved, resulting in
less room for additional sources of antimatter. 

For antiprotons, both the absolute flux and the energy dependence of
the antiproton-to-proton ratio seem to be broadly consistent with a picture
where no dark matter contribution is needed to account for the observations.
The BESS collaboration reported lately a possible excess of antiprotons\,\cite{bessmitchell}
at energies below 1\,GeV, but the effect seems small given the limited 
statistics of the data and the uncertainties related to the proper
description of solar modulation in this energy range. In December 2004, the BESS collaboration
managed for the first time to conduct a 9-day flight in Antarctica that should have doubled the
available antiproton statistics and might give new insight into the
reality of the observed effect.

Positron measurements were conducted as well and analysed in terms of the 
so-called {\em positron fraction} (the fraction of positrons among all
detected electrons and positrons). The energy dependence of the positron fraction has been 
measured between 1 and 30\,GeV, and features of the 
observed spectrum have been interpreted as being due to dark matter.
In particular the data of the HEAT experiment (High Energy Antimatter Telescope)
have been discussed in this context. The HEAT spectrometer has been
flown in two different detector configurations (HEAT-$e^\pm$ (Fig.~\ref{fig:heat} (left)) and
HEAT-pbar). The flights took place near solar maximum (1994 and 1995, HEAT-$e^\pm$) 
and near solar minimum (2000, HEAT-pbar), and at different geomagnetic
cutoffs (around 1\,GeV in 1995, and around 4\,GeV in 1994 and 2000).
The energy dependence of the positron fraction measured during the 1994/95 flights 
agrees with what has been found in 2000\,\cite{heatdata1,heatdata2} (Fig.~\ref{fig:heat} (right)),
and the HEAT collaboration finally concluded that the intensity of positrons
is consistent with a purely secondary origin due to nuclear interactions
in interstellar space\,\cite{heatdata2}.

Given the experimental and model-related uncertainties, a contribution from 
dark matter cannot be ruled out and this possibility has been the 
subject of theoretical studies. 
A popular dark matter candidate motivated by elementary particle physics
(the supersymmetric neutralino) couples in proportion to the mass of
the final state particles and is thus an inefficient electron-positron
source. In this situation, particles expected in Kaluza-Klein theories ofter a viable
alternative and have been studied in detail\,\cite{heathooper} (Fig.~\ref{fig:heat} (right)).  

\section{Electron-Positron Annihilation Radiation}

Another way of tracing positrons is the observation of the 511\,keV 
radiation that is created when electrons and positrons annihilate at rest
somewhere in the Milky Way. 
Energetic positrons suffer rapid ionization losses and can annihilate with 
electrons depending on the atom and electron density in the interstellar medium.
Since the stopping distance of positrons is typically much smaller than their mean
free path for annihilation, positrons do indeed thermalise before
annihilation. The annihilation process itself seems to be dominated
by positronium formation\,\cite{posfrac} creating a narrow line (25\,\% of the 
time) or a three-photon continuum (75\,\% of the time).

Electron-positron annihilation radiation from the Galactic Centre was detected for the first time
in the 70s\,\cite{posdetect,posdetect2,posident}. After initial claims of
a flux variability in time, consistent results for the absolute photon flux and the line shape
were obtained by a number of experiments\,\cite{pos1,pos2,pos3,pos4}.
There was much less agreement on the morphology of the emission region. Some
measurements suggested both a contribution from the bulge and the disk of the Milky
Way, and an additional component from a location at 
positive galactic latitude (i.e.~above the galactic plane) was also 
discussed.

The most recent high-statistics observations of the 511\,keV annihilation
radiation have been obtained using the INTEGRAL satellite. INTEGRAL (INTErnational Gamma-Ray Astrophysics Laboratory)
is an ESA mission in cooperation with Russia and the United States launched in 2002.
One of the instruments aboard INTEGRAL is SPI (SPectrom\`etre Integral),
and this device has been used to measure the flux, line shape and 
morphology of the 511\,keV radiation with improved accuracy. SPI
is a coded-mask telescope with a 16$^\circ$ (FWHM) field of view and BGO collimators. It uses 19 Germanium
detectors to measure the shadow cast by the mask. It is sensitive 
for photon energies between 20\,keV and 10\,MeV and has an energy resolution of 2\,keV (at 1\,MeV).
The coded mask and the pixelation of the Germanium detector result in an angular
resolution around 2$^\circ$.

\begin{figure}
\begin{center}
\begin{tabular}{cc}
\epsfig{figure=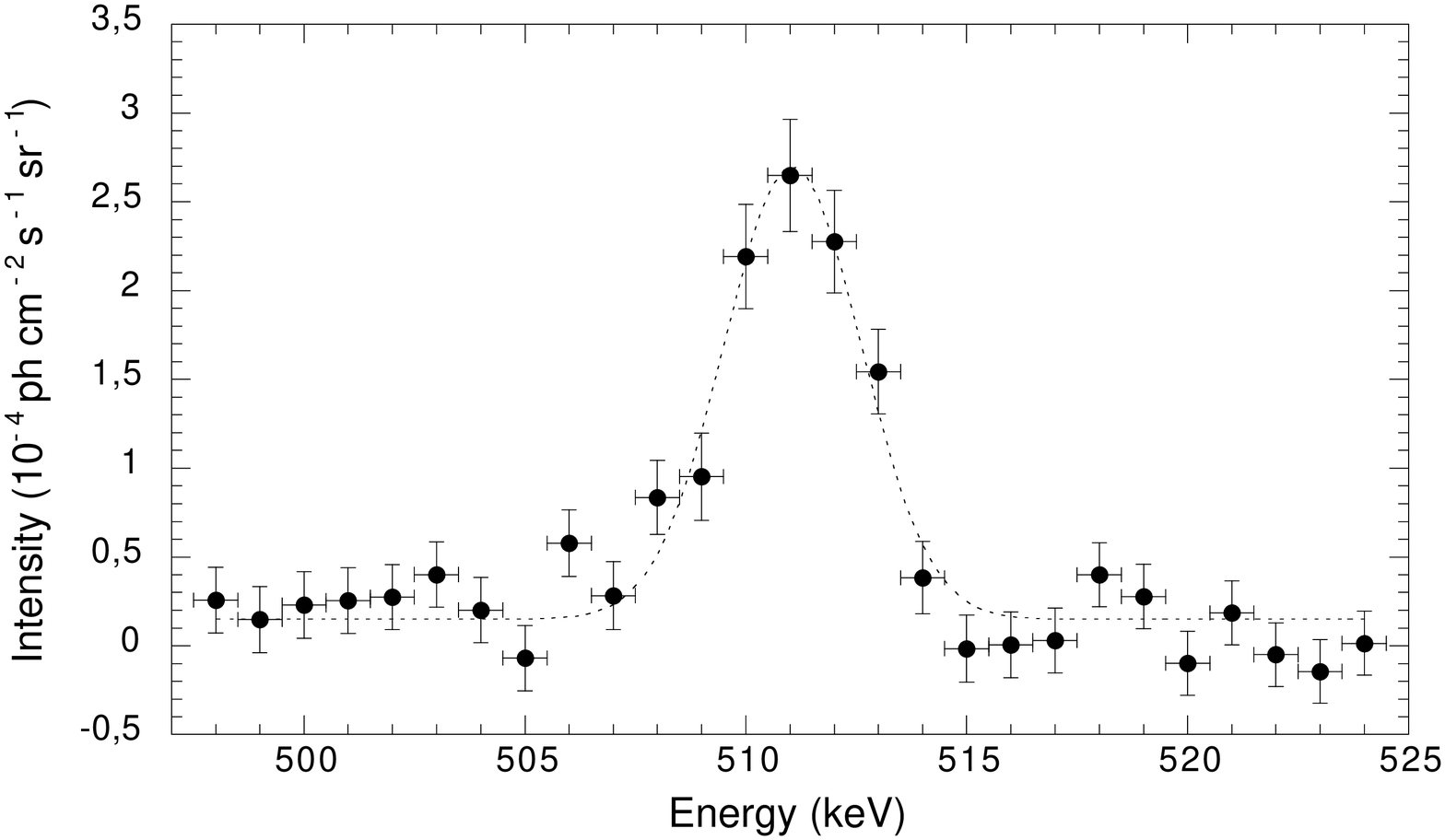,width=0.5\linewidth} &
\psfig{figure=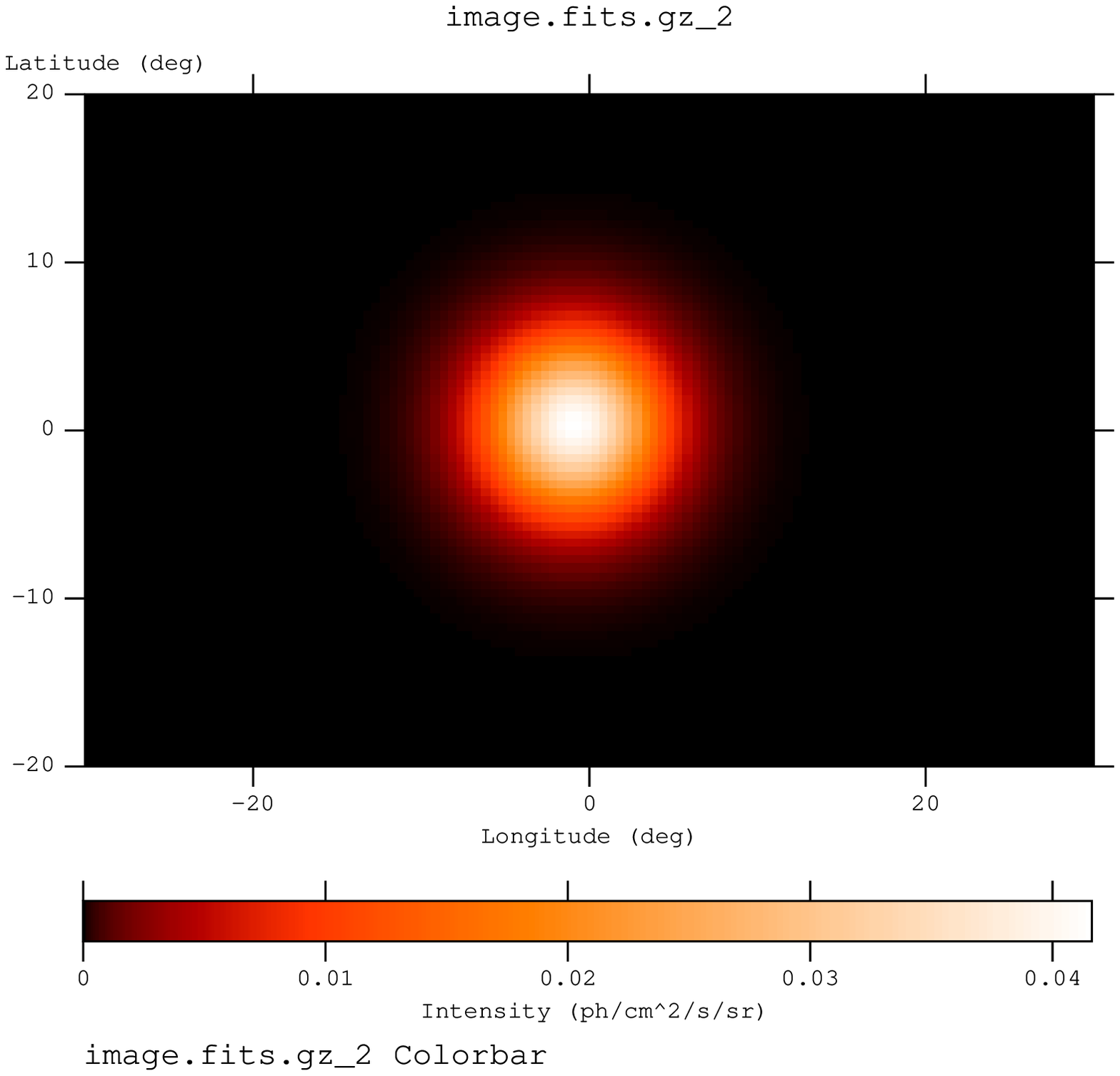,width=0.5\linewidth} 
\end{tabular}
\caption{\label{fig:spi} (Left:) SPI 511\,keV flux spectrum. The dotted line
  is the result of modelling the spectrum using a Gaussian of FWHM of $10^\circ$ 
  around the Galactic Centre\,\protect\cite{spijean}.
  (Right:) Fit model for the 511\,keV line emission from the Galactic Centre
  region. The fit model is a Gaussian representing the Galactic bulge\,\protect\cite{spiweiden}.
}
\end{center}
\end{figure}

The SPI data published so far\,\cite{spijean,spiknoedl,spiweiden} were taken
in 2003 and cover galactic longitudes of $\pm 30^\circ$ and galactic latitudes of $\pm 10^\circ$.
The analysis relies on the accurate subtraction of annihilations initiated by
particles showering up in the detector. The data show a strong signal (Fig.~\ref{fig:spi} (left))
with a significance exceeding 20 standard deviations. The measured flux 
and the line width are consistent with earlier measurements. Despite significant
exposure out to galactic longitudes of $\pm 30^\circ$ the morphology of the emission
region is accurately modelled by a spherical distribution which is centred at the Galactic Centre
and exhibits a Gaussian radial profile with a FWHM of $8^\circ$. Upper limits
on the flux contribution from the galactic disk were derived.

Measurements of the 511\,keV annihilation radiation address 
all kinds of positron sources in the Milky Way. Most of the discussed
sources (like supernovae, pulsars, stellar flares, Wolf-Rayet stars, cosmic-ray interactions
with the interstellar medium and others) would probably imply a non-negligible contribution 
from the galactic disk. The SPI measurement suggests the galactic bulge as the positron source
and this agrees with the notion that the density of dark matter particles should
be highest in the Galactic Centre region, while a contribution from the galactic disk
is not necessarily expected. This has resulted in attempts to model the SPI data as being
due to dark matter annihilation and we refer the reader to a contribution 
in these proceedings\,\cite{boehm} which discusses this possibility in detail.

\section{Diffuse $\gamma$-ray Emission from the Milky Way}

\begin{figure}
\begin{center}
\begin{tabular}{cc}
\epsfig{figure=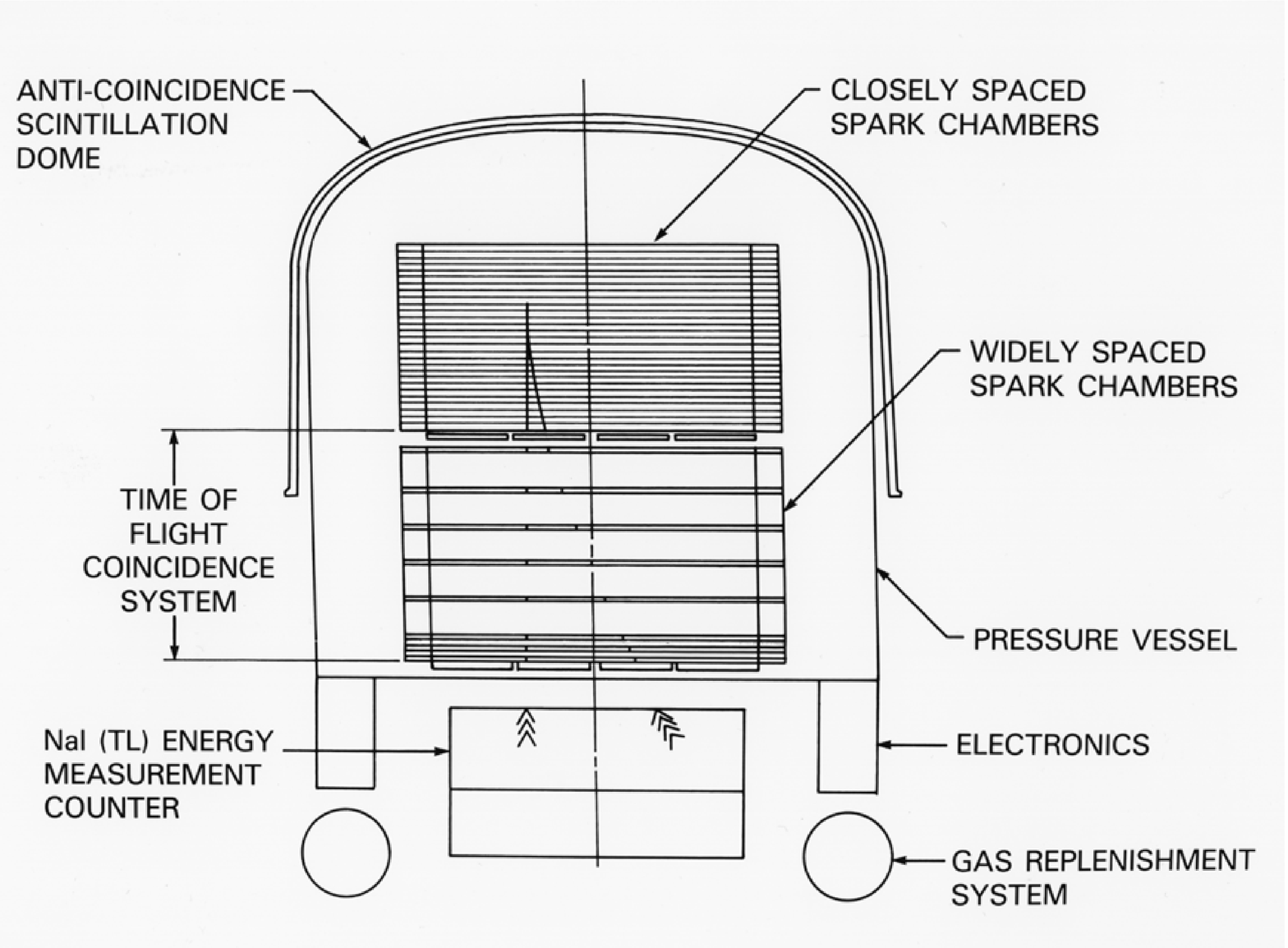,width=0.5\linewidth} &
\epsfig{figure=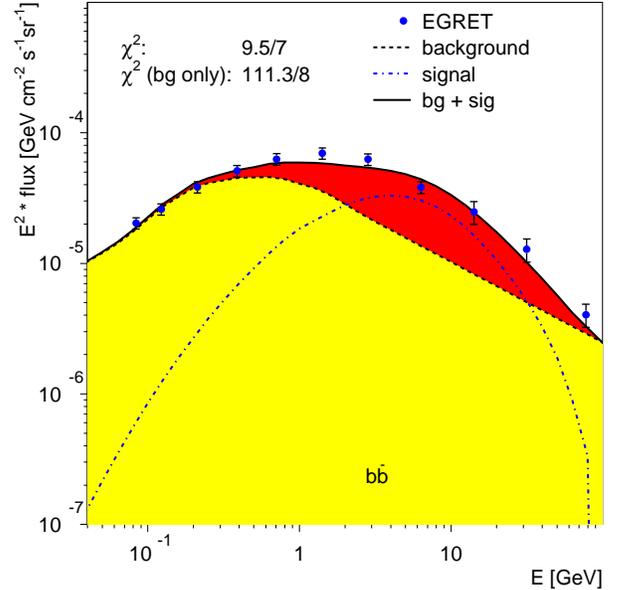,width=0.5\linewidth} 
\end{tabular}
\caption{\label{fig:egret} (Left:) Cross section of the EGRET detector 
  aboard CGRO. (Right:) Diffuse $\gamma$-ray energy spectrum of the Inner Galaxy
  (longitude $330-30^\circ$, latitude $|b|<5^\circ$). The EGRET data are shown as dots.
  The dashed line denotes the photon contribution from hadronic interactions
  in the interstellar gas, the dot-dashed line denotes the photon spectrum
  expected from neutralino annihilations into $b\bar{b}$ pairs. The solid line
  is the sum of the two components\,\protect\cite{wimdeboer}.
}
\end{center}
\end{figure}

Photons from the decay cascade of particles created by annihilating dark matter
can potentially make a significant contribution to the diffuse $\gamma$-ray emission
from the Milky Way and this possibility has lately been investigated\,\cite{wimdeboer}
using EGRET data. The EGRET (Energetic Gamma Ray Experiment Telescope) instrument is
a detector sensitive to photons with energies between 20\,MeV and 30\,GeV that was
flown between 1991 and 2000 aboard NASA's Compton Gamma-Ray Observatory (CGRO). 
EGRET detected photons by conversion into 
electron-positron pairs and subsequent measurement of the tracks and 
the particle energies using a spark chamber and a calorimeter (Fig.~\ref{fig:egret} (left)).
The photon energy was reconstructed with an accuracy of 20\,\%, and the 
angular resolution varied between 1.3$^\circ$ (at 1\,GeV) and 0.4$^\circ$ (at 10\,GeV).

The EGRET observations discovered several hundred $\gamma$-ray sources. Diffuse
emission was searched for by subtracting those sources from the overall photon
flux, and it was noted already in 1997 that the energy spectra of the diffuse 
component show a 60\,\% excess compared with model predictions\,\cite{hunter}. 
This excess was found at energies above several GeV, while the EGRET data
below 1\,GeV were well accounted for by models that describe the photon spectrum
as entirely due to secondary interactions of charged cosmic rays. The excess
was found with a very similar spectral shape for all sky directions, i.e.\ it 
does not only occur when observing, for example, the Galactic Centre region.

An excess of $\gamma$-rays from all directions is a signature that could 
be provided by dark matter annihilations taking place throughout the Milky
Way. The absolute photon flux from a certain direction is proportional 
to the line-of-sight integral over the squared dark matter density, since
the annihilation rate goes with the density squared. The shape of the 
energy spectrum of the excess photons is determined by the mass and 
the couplings of the dark matter particle and should indeed be independent of
the observation direction. An analysis of the total EGRET data set has been 
performed to investigate possible dark matter contributions to the 
diffuse photon flux\,\cite{wimdeboer}, and is described in detail elsewhere in
these proceedings\,\cite{wimproceed}. The analysis extracts the background 
shape from a model of cosmic ray propagation and $\gamma$-ray 
production in the Galaxy (GALPROP\,\cite{model2})
with a particular sets of parameters, while the signal shape is based
on expectations from supersymmetry\,\cite{darksusy} combined with the known 
fragmentation of quarks. The absolute normalizations of signal and 
background are allowed to vary, and a fit of the energy spectra is used
to limit the mass of the dark matter particle to the 50--100\,GeV range (Fig.~\ref{fig:egret} (right)).
The spatial distribution of the detected photons is used to extract
very detailed information about the dark matter distribution in the Galaxy.
It is found that the EGRET data can be described with a contribution
from annihilating sypersymmetric dark matter when the annihilation rate
is enhanced by a 'boost factor' of 20 which could be possible due to clumping on small
scales.

The proper modelling of the sizable fraction of background photons 
created by interactions of charged cosmic rays is a key element when
looking for dark matter induced photons. The GALPROP model combined with  
cosmic ray measurements (i.e.\ local (anti)proton and electron spectra) 
has been used to describe the EGRET data on diffuse photon emission 
without any dark matter contribution. This alternative analysis\,\cite{smreimer} 
uses the diffuse emission itself to adjust the electron injection spectrum
and also scales up the initial proton spectum using antiproton data. 
It arrives at a different GALPROP parameter set and somewhat relaxes the 
requirement on the agreement of locally measured and predicted 
electron and proton spectra. This is not implausible since details
of the galactic structure are anyway not taken into account in the
background model. 

\section{Very High-Energy $\gamma$-rays from the Galactic Centre}

The Galactic Centre region is considered as a promising target for
dark matter searches since an increased dark matter density is 
expected there. The dynamical centre of the Galaxy is dominated 
by Sagittarius $A^\ast$, a putative black hole of $\approx 3\,\cdot 10^6$ solar masses\,\cite{schoedel}.
Sgr $A^\ast$ was first discovered as compact radio source and
exhibits a very low bolometric luminosity. It was found to be 
variable on sub-hour timescales in the infrared band and in X-rays\,\cite{genzel,x1,x2,x3} 
suggesting that the observed radiation is created close 
to the event horizon of the black hole. Sgr $A^\ast$ is surrounded
by a supernova remnant (Sgr $A$ East) and a region with a high density
of ionized hydrogen (Sgr $A$ West).

\begin{figure}
\begin{center}
\begin{tabular}{cc}
\epsfig{figure=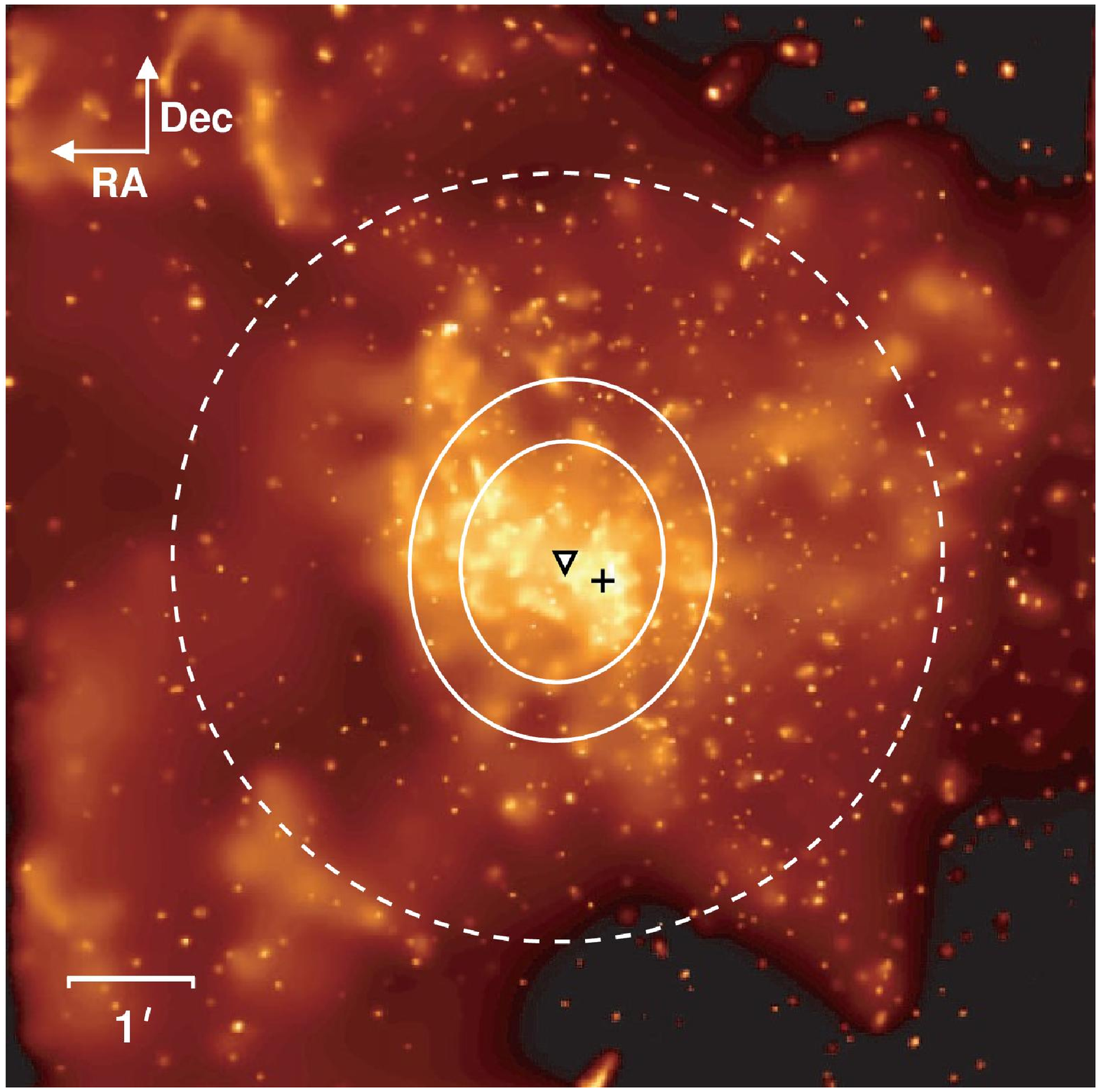,width=0.5\linewidth} &
\epsfig{figure=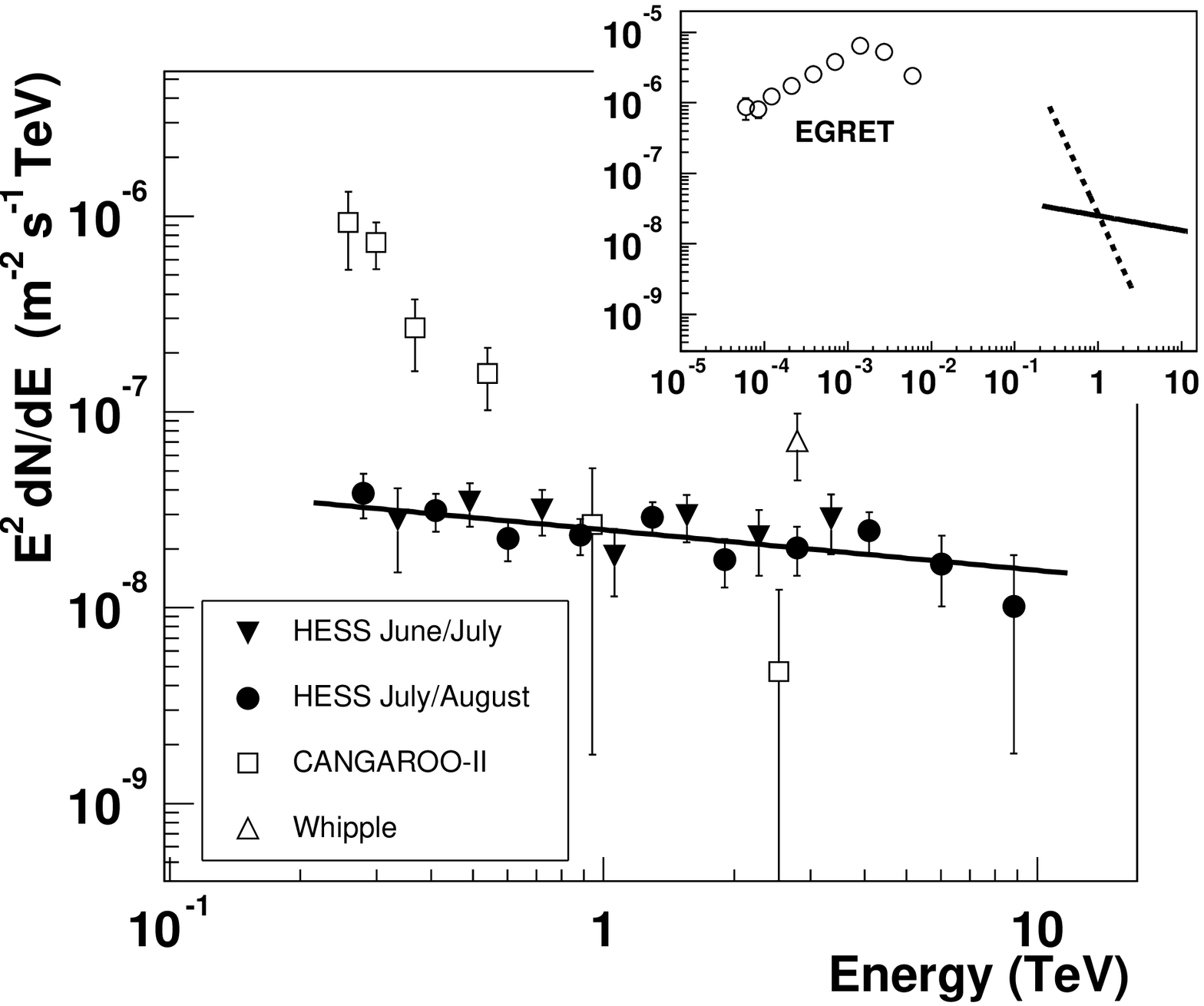,width=0.5\linewidth} 
\end{tabular}
\caption{(Left:) Location of the H.E.S.S.\ VHE $\gamma$-ray signal shown 
  on a $8.5'$ by $8.5'$ Chandra X-ray map. The triangle denotes the centre
  of gravity of the H.E.S.S.\ signal, the cross marks the location of Sgr $A^\ast$.
  The contour lines indicate the 68\,\% and 95\,
  position; the dashed line is the upper limit on the rms source size (95\,\% confidence level).
  (Right:) Energy spectrum of the $\gamma$-rays from the Galactic Centre. 
  The full circles and triangles denote H.E.S.S.\ data taken in 2003 with different
  detector configurations. The open squares and the open triangle correspond
  to detections of the galactic centre region by CANGAROO\,\protect\cite{gccangaroo} (2001/2002)
  and Whipple\,\protect\cite{gcwhipple} (1995--2003).
}
\label{fig:hess}
\end{center}
\end{figure}

In 2004, results of observations of the Galactic Centre in very high-energy $\gamma$-rays ($>100\,\mbox{GeV}$) 
have been published by several experiments using the imaging atmospheric Cherenkov technique.
This technique records images of air showers by focussing
the Cherenkov light emitted by the secondary shower particles onto a camera.
Energy, direction and type ($\gamma$-ray or background hadron) of the particle that
initiated the air shower can be extracted from an analysis of the camera image.
Viewing the same shower with more than one telescope allows to reconstruct the 
direction of the primary with a precision of 0.1$^\circ$ per event; 
the energy is estimated with a typical relative resolution of 20\,\%.

Detections of very high-energy $\gamma$-rays from the Galactic Centre have 
been published by the Whipple, CANGAROO and H.E.S.S.\ groups\,\cite{gcwhipple,gccangaroo,gchess}.  
The most significant signal (11\,$\sigma$) was reported by the H.E.S.S.\ experiment
using data from two telescopes in 2003. H.E.S.S.\ is located in Nambia and can observe the Galactic Centre 
under small zenith angles resulting in an energy threshold of 160\,GeV for this data set.
The detected $\gamma$-ray source appears point-like compared with the
point spread function of the instrument. The H.E.S.S.\ measurement 
pinpoints the source position with a much higher accuracy than the earlier
Whipple and CANGAROO detections. Within errors, the source position is compatible
with Sgr $A^\ast$, but also with the supernova remnant Sgr $A$ East (Fig.~\ref{fig:hess} (left)). 
The energy spectrum measured by H.E.S.S.\ is well described by a power law $E^{-\Gamma}$
with a spectral index $\Gamma$ of 2.2 (full symbols in Fig.~\ref{fig:hess} (right)). 
This spectrum is significantly flatter than the spectrum reported by CANGAROO (open squares in Fig.~\ref{fig:hess} (right)).
The CANGAROO spectrum combines data from 2001 and 2002 and implies
a higher overall photon flux than the H.E.S.S. measurement. 

The different flux levels and spectral shapes reported by the three experiments
seem to suggest a source variability in time. This option appears however unlikely since
none of the individual groups found indications for variability in its data set.
Simultaneous observations and a cross-calibration of experiments with known
steady sources should allow more insights into this problem. The H.E.S.S.\ detection
has lately been confirmed by data taken in 2004 with four telescopes. The 
overall significance of the signal is now at the 35\,$\sigma$ level, and both
source position and shape of the energy spectrum agree completely with what
has been found in 2003. The spectrum has been extended to energies as low
as 200\,GeV and as high as 20\,TeV, and is still well described by a power law.

The $\gamma$-ray signal from the Galactic Centre has been interpreted 
both as dark matter\,\cite{gchooper} and in terms of 'ordinary' astrophysical processes.
Ascribing the total photon signal to dark matter annihilations and
fitting the H.E.S.S.\ energy spectrum to a shape expected from 
a weakly interacting supersymmetric particle\,\cite{gchorns}, results in a lower limit
of 12\,TeV on the mass of the dark matter particle. The mass limit is driven to this high value since the 
H.E.S.S.\ energy spectrum shows no evidence for a cutoff at higher energies. 
Also, the required annihilation cross section is rather high,
even though the angular distribution of the signal is already compatible 
with a steep increase of the dark matter density towards the Galactic Centre.
Other interpretations of the Galactic Centre signal invoke Kaluza-Klein dark 
matter\,\cite{bergstroem} and find that the entire H.E.S.S.\ data can be 
described with a particle mass of roughly 10\,TeV for boost factors of the
order $10^3$. All these interpretations are certainly pushed to their
limits by the 2004 H.E.S.S.\ data which extend with a pure power law
shape to even higher energies. At the same time, explanations are being
followed where only a fraction of the detected photon flux is 
attributed to dark matter.

Astrophysical explanations of the Galactic Centre signal have been
studied as well and could be viable alternatives. Accretion onto the black hole Sgr $A^\ast$ 
and associated acceleration processes can produce high-energy protons and electrons
that create secondary $\gamma$-rays due to synchrotron and curvature radiation
or due to interactions with ambient matter and photon fields. The low 
luminosity of Sgr $A^\ast$ implies that it appears transparent for TeV $\gamma$-rays,
and it has been demonstrated\,\cite{aha} that the H.E.S.S.\ data can be 
explained assuming populations of high-energy protons or electrons
associated with Sgr $A^\ast$. More detailed and simultaneous studies 
in different wavelength bands will help to tackle the question whether
the very-high energy $\gamma$ radiation is indeed related to Sgr $A^\ast$. 

Other astrophysical interpretations focus on the supernova
remnant Sgr $A$ East\,\cite{crocker}. The spectral index measured by H.E.S.S.\ is close
to the value expected for particle acceleration in the shock front of 
supernova explosions. The observed $\gamma$-rays would then be 
created by protons and/or electrons accelerated to energies of
${\cal O}(100\,\mbox{TeV})$. This explanation appears not unmotivated 
since Sgr $A$ East is a very powerful supernova remnant with an
above-average explosion energy, and the presence of a 100\,TeV-particles 
has lately been demonstrated for other supernova remnants\,\cite{rxj,velajr}. 

\section{Summary and Outlook}

A number of signals recorded by balloon experiments, satellites and imaging 
Cherenkov telescopes have been interpreted as being due to dark matter. 
Dark matter explanations, often of the same signal, invoke various theoretical 
frameworks, and there seems to be no scenario that can explain more than 
experimental signature. In most cases, astrophysical sources of the 
observed signals offer a reasonable alternative. In absence of an unambiguous dark 
matter signal (like a line emission), indirect searches are made difficult
by uncertainties related to the dark matter distribution in the Galaxy and the 
presence of various background processes. In particular a more precise
modelling of the sources, spectra and propagation of ordinary charged 
cosmic rays is required to predict the shape of backgrounds with greater
confidence. Long-term studies of charged particle spectra and of the impact of solar
modulation using future space-based experiments (like AMS-02 and PAMELA\,\cite{ams,pamela})
will provide valuable information in that respect, and will detect
antiprotons and positrons at the same time. Imaging atmospheric Cherenkov telescopes
are exploring the skies with much-improved sensitivity and continue to
discover sources some of which seem to have no counterpart in any other
waveband\,\cite{scan}. They might also encounter dark matter clumps that
are predicted to be bright $\gamma$-ray sources based on simulation of
cosmological models\,\cite{benmoore}. These searches will certainly 
benefit from the overlap towards lower photon energies ($<100\,\mbox{GeV}$)
provided by the future GLAST satellite mission\,\cite{glast}.

\section*{Acknowledgments}

The author would like to thank O.\ Reimer and J.\ Ripken for discussions related to
the material presented here.


\section*{References}
\begin{thebibliography}{99}
\bibitem{review} G. Bertone, D. Hooper and J. Silk, hep-ph/0404175.
\bibitem{model1} V.~S.~Ptuskin et al., A\&A {\bf 321}, 434 (1997).
\bibitem{model2} A.~W.~Strong \& I.~V.~Moskalenko, ApJ {\bf 509}, 212 (1998).

\bibitem{bessmitchell}J.~W.~Mitchell et al., Advances In Space Research (2005).
\bibitem{heatdata1} S.~W.~Barwick et al., ApJ {\bf 482}, L191--194 (1997).
\bibitem{heatdata2} M.~A.~DuVernois et al., ApJ {\bf 559}, L296--303 (2001).
\bibitem{heathooper} D.~Hooper, hep-ex/0409272).

\bibitem{posfrac} R.~Kinzer et al., ApJ {\bf 559}, 282 (2001).
\bibitem{posdetect} W.~N.~Johnson et al., ApJ {\bf 172}, L1 (1972).
\bibitem{posdetect2} W.~N.~Johnson \& R.~C.~Haymes, ApJ {\bf 184}, 103 (1973).
\bibitem{posident} M.~Leventhal et al., ApJ {\bf 225}, L11 (1978).

\bibitem{pos1} W.~A.~Mahoney et al., ApJ Supp.~Ser. {\bf 92}, 387 (1994).
\bibitem{pos2} M.~Leventhal et al., ApJ {\bf 405}, L25 (1993).
\bibitem{pos3} D.~M.~Smith et al., ApJ {\bf 414}, 165 (1993).
\bibitem{pos4} M.~J.~Harris et al., ApJ {\bf 501}, L55 (1998).

\bibitem{spijean} P.~Jean et al., A\&A {\bf 407}, L55--L58 (2003).
\bibitem{spiknoedl} J.~Kn\"odlseder et al., A\&A {\bf 411}, L457--L460 (2003).
\bibitem{spiweiden} G.~Weidenspointner et al., 
Proc.\ of the V INTEGRAL Workshop, Munich 16--20 Feb 2004 
[astro-ph/0406178].
\bibitem{boehm} C.~Boehm, these proceedings.

\bibitem{wimdeboer} W.~de~Boer et al., astro-ph/0408272.
\bibitem{hunter} S.~D.~Hunter et al., ApJ {\bf 482}, 205 (1997).
\bibitem{wimproceed} W.~de~Boer, these proceedings.
\bibitem{darksusy} P.~Gondolo et al., JCAP 0407 (2004) [astro-ph/0406204].
\bibitem{smreimer} A.~W.~Strong et al., ApJ {\bf 613}, 956 (2004).

\bibitem{schoedel} R.~Sch\"odel et al, Nature {\bf 419}, 694 (2002).
\bibitem{genzel} R.~Genzel et al, Nature {\bf 425}, 934 (2003).
\bibitem{x1} F.~K.~Baganoff et al., Nature {\bf 413}, 45 (2001).
\bibitem{x2} A.~Goldwurm et al., ApJ {\bf 584}, 751 (2003).
\bibitem{x3} D.~Porquet et al., A\&A {\bf 407}, L17 (2003).
\bibitem{gcwhipple} K.~Kosack et al., ApJ {\bf 608}, L97 (2004).
\bibitem{gccangaroo} K.~Tsuchiya et al., ApJ {\bf 606}, L115 (2004).
\bibitem{gchess} F.~Aharonian et al. [The H.E.S.S.~Collaboration], A\&A {\bf 425}, L13--L17 (2004).
\bibitem{gchooper} D.~Hooper et al., astro-ph/0404205.
\bibitem{gchorns} D.~Horns, astro-ph/0408192. 
\bibitem{bergstroem} L.~Bergstr\"om et al., astro-ph/0410359.
\bibitem{aha} F.~Aharonian \& A.~Neronov, astro-ph/0408303.
\bibitem{crocker} R.~M.~Crocker et al., astro-ph/0408183.
\bibitem{rxj} F.~Aharonian et al. [The H.E.S.S.~Collaboration], Nature {\bf 432}, 74 (2004).
\bibitem{velajr} M.~Lemoine-Goumard, Proc.~of XL$^{\mbox{th}}$ Moriond Conf.~``Very High Energy Phenomena in the Universe'' (2005).

\bibitem{ams} http://ams.cern.ch/AMS
\bibitem{pamela} M.~Circella, NIM A {\bf 518}, 153 (2004).
\bibitem{scan} F.~Aharonian et al. [The H.E.S.S.~Collaboration], Science {\bf 307}, 1938 (2005).
\bibitem{benmoore} J.~Diemand et al., Nature {\bf 433}, 389 (2005).
\bibitem{glast} http://www-glast.stanford.edu

\end{thebibliography}
\end{document}